\def\beq{\begin{equation}}
\def\eeq{\end{equation}}

\def\beqn{\begin{eqnarray}}
\def\eeqn{\end{eqnarray}}

\newcommand\sss{\scriptscriptstyle\rm}
\newcommand\as{\alpha_{\rm s}}
\newcommand\MCatNLO{{\rm MC}@{\rm NLO}}
\newcommand\stepf{\Theta}
\newcommand\xm{x_{\sss M}}
\newcommand\IMC{I_{\sss MC}}
\newcommand\xsecO{\frac{d\sigma}{dO}}
\newcommand\orda{{\cal O}(a)}
\def\VEV#1{\left\langle #1\right\rangle}

\documentclass{camera-mod}

\begin{document}

%
\title{MATCHING NLO COMPUTATIONS AND PARTON SHOWERS IN QCD
\footnote{Talk at XIV IFAE, 3--5 April 2002, Parma, Italy.}}

%
\author{Stefano Frixione}

\organization{Laboratoire d'Annecy-le-Vieux de Physique de Particules\\
  Chemin de Bellevue, BP 110,
  74941 Annecy-le-Vieux CEDEX, France~%
\footnote{On leave of absence from INFN, Sez. di Genova, Italy.}}

\maketitle

Reliable predictions of cross sections and final-state distributions
for QCD processes are a crucial ingredient in high-energy collider
experiments, not only as a test of QCD but also for new particle
searches.  All systematic approaches to this problem are based on
fixed-order (FO) results in perturbation theory, and yield (usually at
the next-to-leading order, NLO) the best available results for
sufficiently inclusive observables. However, in many cases a more
exclusive description of final states is needed. In such cases, in
which one also combines perturbative calculations with a model
for the conversion of partonic final states into hadrons, 
Monte Carlo (MC) simulations are generally adopted. MC's
operate on partonic states with high multiplicity and low relative
transverse momenta, which are obtained from a parton shower or dipole
cascade approximation to QCD dynamics.  This has to be confronted with
FO results, which can describe the complementary region of small
multiplicities, and large relative transverse momenta.

The lack of large transverse momentum emissions, and the fact that total 
rates are computed to leading order accuracy only, are serious problems in
MC simulations, especially when the CM energies are in the TeV range. 
These problems 
can be solved by a suitable combination of MC and FO methods. Given the 
flexibility of MC's, it is actually desirable to use some FO techniques
in the framework of MC simulations. One approach (matrix element
corrections) acts at the level of the generation of hard processes in 
the MC: $2\to n$ processes, with $n>2$, are given to the parton shower
as initial conditions (in standard MC's, hard processes have usually a 
$2\to 2$ kinematics). ME corrections therefore improve the description 
of large transverse momentum emissions, but results for total rates have
still a LO accuracy.

A more involved approach to the problem aims at improving also the 
computation of total rates. There is
a considerable freedom in the very definition of such an approach.
Here I shall follow the formalism presented in ref.~\cite{Frixione:2002ik}, 
denoted by $\MCatNLO$ there. A $\MCatNLO$ is required to have the
following features: {\em a)}~The output is a set of events, which are
fully exclusive. {\em b)}~Total rates are accurate to NLO. {\em c)}~%
NLO results for all observables are recovered upon expansion of
$\MCatNLO$ results in $\as$. {\em d)}~Hard emissions are treated as in
NLO computations. {\em e)}~Soft/collinear emissions are treated as in
MC. {\em f)}~The matching between hard- and soft-emission regions is
smooth. {\em g)}~MC hadronization models are adopted. Condition
{\em c)} is the requirement that the $\MCatNLO$ be not affected by double
counting problems. In ME corrections, double counting corresponds to
generating the same kinematical configuration twice (by hard process
generation, and by means of the shower). On the other hand, the 
presence of negative weights in the $\MCatNLO$ implies that a double
counting can be due both to an excess and to a lack of emission.

The implementation of the $\MCatNLO$ formalism is technically complicated
in QCD. However, it is possible to understand the basic features of the
findings of ref.~\cite{Frixione:2002ik} by studying a toy model. In this
model, a system can radiate massless ``photons'', whose energy I denote 
by $x$, with $0\le x \le x_s\le 1$, $x_s$ being the energy of the system
before the radiation. After the radiation, the energy of the system is
$x_s^\prime=x_s-x$. The system can undergo several further emissions; 
on the other hand, one photon cannot split further. The NLO predictions
for the mean value of an observable $O$ (possibly including $\stepf$
functions that define an histogram bin) is
\beq
\VEV{O}=\int_0^1 dx \left[O(x)\frac{aR(x)}{x}
+O(0)\left(B+aV-\frac{aB}{x}\right)\right],
\label{nlosubtint}
\eeq
where $a$ is the coupling constant, and $B$, $V$, and $R(x)$ are the
Born, (finite) virtual, and real contributions to the cross section 
respectively. The function $O(x)$ returns the value of the observable
$O$ in the cases of real photon emission ($x\ne 0$), and of virtual photon
emission or of no emission ($x=0$). Eq.~(\ref{nlosubtint}) is based on 
standard subtraction techniques for the cancellation of infrared
divergences. In the toy model, a MC can be defined with 
the following Sudakov form factor
\beq
\Delta(x_1,x_2)=\exp\left[-a\int_{x_1}^{x_2}dz\frac{Q(z)}{z}\right],
\label{Deltadef}
\eeq
where $Q(z)$ is a monotonic function, such that $0\le Q(z)\le 1$, 
and $\lim_{z\to 0}Q(z)=1$. The form factor in 
eq.~(\ref{Deltadef}) can be used in a standard way to generate showers
in which the system can emit any number of photons. Since this number
is in general larger than 1, the function $O(x)$ is not sufficient to
describe the observable $O$; we denote the distribution in the observable 
$O$ obtained from the MC by
\beq
\IMC(O,\xm(x))\,,
\label{IMCdef}
\eeq
where $\xm(x)$ indicates the initial condition for the shower (the
system has energy $1-x$).

A naive attempt at matching NLO computations and MC simulations amounts
to formally replacing $O(x)$ with $\IMC(O,\xm(x))$ in eq.~(\ref{nlosubtint}):
\beq
\left(\xsecO\right)_{\sss naive}
=\int_0^1 dx \Bigg[\IMC(O,\xm(x))\frac{aR(x)}{x}
+\IMC(O,1)\left(B+aV-\frac{aB}{x}\right)\Bigg].
\label{IMCnlonaive}
\eeq
Here, a couple of MC runs are performed, one in which the initial condition
is given by the system plus a photon of energy $x$ ($\IMC(O,\xm(x))$),
and one in which there is no photon emission prior to the shower
($\IMC(O,1)$); the results are then weighted with $aR(x)/x$ and
$B+aV-aB/x$ respectively, as in eq.~(\ref{nlosubtint}). Unfortunately,
this does not work: the procedure in eq.~(\ref{IMCnlonaive}) does have
double counting. The solution proposed in ref.~\cite{Frixione:2002ik}
(called {\em modified subtraction} there) is the following:
\beqn
\left(\xsecO\right)_{\sss msub}
&=&\int_0^1 dx \Bigg[\IMC(O,\xm(x))\frac{a[R(x)-BQ(x)]}{x}
\nonumber \\*
&&+\IMC(O,1)\left(B+aV+\frac{aB[Q(x)-1]}{x}\right)\Bigg]\;,
\label{IMCfive}
\eeqn
which is obtained from eq.~(\ref{IMCnlonaive}) by subtracting and 
adding the quantity $\IMC(O,\xm)\,aBQ(x)/x$, using $\xm=\xm(x)$ in 
the first and $\xm=1$ in the second term introduced in this way. 
These two terms compensate, at $\orda$, the shower contribution due 
to $B\IMC(O,1)$, and therefore double counting does not occur. 
This happens because these terms coincide with the $\orda$
term in the expansion of the Sudakov~(\ref{Deltadef}), the two
different kinematics ($\xm(x)$ and 1) accounting for the emission 
probability and for 
the no-branching probability. Furthermore, the weights
appearing in eq.~(\ref{IMCfive}) are finite (this is not the 
case in eq.~(\ref{IMCnlonaive})), and this allows efficient
unweighting.

In order to deal with the case of QCD, one has to formally
translate eq.~(\ref{IMCfive}). This is essentially a technical task;
however, there are a couple conceptual issues which do not arise in
the simple toy model. In QCD, there are two types of non-UV singularities
(soft and collinear), whereas in the toy model only soft singularities
appear. In the case of collinear emission from initial-state partons, 
the resulting kinematics is one in which one massless parton is 
going down the beam line with non-zero energy. Such a configuration
cannot be used as an initial condition for a parton shower; however,
one can always boost the system to a frame in which the emitted parton
has zero energy. MC-wise, this procedure can be made efficient by a
suitable redefinition of the Bjorken $x$'s~\cite{Frixione:2002ik}.
The second problem is that, in the soft limit, 
the ${\cal O}(\as)$ term in the expansion
of the MC result does not have the angular distribution expected from
QCD factorization theorems, and this prevents a straightforward use
of this result in the construction of the $\MCatNLO$. However, the 
total amount of soft energy radiated by the MC is in agreement with
QCD expectation, and this is sufficient for any infrared-safe observable
to be predicted correctly.

A public version of the $\MCatNLO$ advocated in 
ref.~\cite{Frixione:2002ik} 
has been presented in ref.~\cite{Frixione:2002bd}. The production of
standard model vector boson pairs ($ZZ$, $W^+W^-$, and $W^\pm Z$)
in hadronic collisions has been implemented, matching the relevant
NLO computations with Herwig. The code can be downloaded from
{\tt http://www.hep.phy.cam.ac.uk/theory/webber/MCatNLO}. The
implementation of the hadroproduction of heavy quarks in
under way~\cite{fnw}. Apart from its phenomenological relevance, it
has to be stressed that the solution of the {\em technical} problems
in the implementation of this process will make the implementation 
of any other production process in the $\MCatNLO$ a straightforward task.

It is a pleasure to thank B. Webber for 
numerous discussions on this and other matters.


\begin{thebibliography}{100}

\bibitem{Frixione:2002ik}
S.~Frixione and B.~R.~Webber,
JHEP {\bf 0206} (2002) 029.

\bibitem{Frixione:2002bd}
S.~Frixione and B.~R.~Webber,
arXiv:hep-ph/0207182.

\bibitem{fnw}
S.~Frixione, P.~Nason, and B.~R.~Webber, in preparation.

\end{thebibliography}
\end{document}